# Complementary Fractional Dimensional Order of Nyquist Sinc Sequences for Time Division Multiplexing

Ali Dorostkar

Isfahan Mathematics House, Isfahan, Iran.
(Email: m110alidorostkar@gmail.com)

**Abstract**: High speed data transmission is enabled by time and wavelength division multiplexing. Here is introduce fractional dimension order of Nyquist pulses sequences for orthogonal time division multiplexing. Firstly, with a representation of the Nyquist sinc sequence by a cosine Fourier series, in one side it is introduced the complementary Nyquist sinc sequences as a better option for data transmission. The results show that data transmission with complementary Nyquist sinc sequences has a better performance in bit error rate for higher signal to noise ratio. On the other side, a possibility of optical time delay by an electrical phase shifter for optical time division multiplexing is theoretically demonstrated. The method can be used in optical time division multiplexing without requiring any optical line delay, granting simple tunability and time shift of Nyquist pulses. In continue, the fractional dimensional order of signal is defined to open a new window for data transmission. Moving of function in fractional dimension can be realized as a new freedom for a signal processing. In other words, dimension itself is a dimension. For a signal of f(t), beside the variation of function in time domain, intrinsically function also has a variation in dimension. In principle, dimension is coupled to every independent variables of function. In this regard, dimensional transformation is introduced to give a mapping of the function in dimensional domain or variations of function in fractional dimension. This mapping gives more information about signal in different point of view like Fourier transformation. Then, the complementary fractional dimensional order of Nyquist sinc sequences is defined to reach higher data rate. It has not a unique solution, however; the best set of solutions for data transmission must be taken in to accounted. Furthermore, the trajectory of fractional dimension can be found by a numerical iterative algorithm which will be explained in the appendix.

## Part I. Conventional & Complementary Nyquist Sinc Sequences

### I. INTRODUCTION

According to the Cisco Visual Networking Index Forecast the overall internet traffic will grow at a compound annual growth rate of 24% from 2016 to 2021 [1]. Especially new applications like video streaming, gaming or cloud computing for big data lead to an ever increasing capacity demand in the communication and inter-data-center networks [2]. The current trend suggests that the demand for higher capacity in high-performance computing, internet data centres and carrier central office applications would reach 1Tbit/s and beyond in the near future [3]. Like optical orthogonal frequency division multiplexing (O-OFDM), transmission based on Nyquist pulses enables an ideal spectral efficiency of 1 symbol/s/Hz. However, Nyquist transmission offers many unique advantages compared to O-OFDM: it requires lower receiver and transmitter bandwidths and accordingly lower speed of analog-to-digital converters [4]. Thus, it is an ideal candidate for a possible integration in silicon photonics. For higher data rate, Nyquist orthogonal time division multiplexing (OTDM) can be combined with a wavelength division multiplexing (WDM) system. WDM tributaries can be transmitted and processed asynchronously [5], a property that is especially important in data center applications. Nyquist pulse shaping leads to lower peak-to-average power ratios [6]. Thus, these pulses are less sensitive to fiber nonlinearities [7] which makes them attractive for long haul transmission systems as well.

However, for the exploitation of the whole available bandwidth of an optical fiber, the Nyquist WDM tributaries must be multiplexed without any guard band. Thus, in order to multiplex the channels, the single channel must have a rectangular bandwidth.

Nyquist pulses possess the special property of zero-inter-symbol interference. The amplitude waveform of Nyquist pulses can be expressed in the time domain as [8-9].

$$f(t) = \frac{\sin(\pi t / \Delta T)}{\pi t / \Delta T} \times \frac{\cos(\alpha \pi t / \Delta T)}{1 - (\alpha t / \Delta T)^2} \qquad (1)$$

Where $\Delta T$ is the pulse duration from the maximum to the first zero crossing and $\alpha$ is known as a roll-off factor [8], which is in the range $0 \leq \alpha \leq 1$. Only the sinc shaped pulse with $\alpha = 0$ has a perfect rectangular spectrum, which possibly leads to a multiplexing of the channels without any guard band. But, the sinc pulse is infinite and hence, just a mathematical construct.

Recently, a Nyquist sinc sequence (NSS) was presented with exception quality and an almost rectangular spectrum [10]. By cascading of intensity modulator with an appropriate control of RF signals and DC bias voltages, unlimited superposition of time-shifted sinc pulses can be achieved, and can be expressed by [10]:

$$\sum_{n=-\infty}^{+\infty} \mathrm{sinc}\left(N\Delta f\left(t - \frac{n}{\Delta f}\right)\right) = \frac{\sin(N\pi\Delta ft)}{N\sin(\pi\Delta ft)} \qquad (2)$$

Where $N\Delta f$ is the total bandwidth of the pulses with a repetition rate of $\Delta f$. In the frequency domain this corresponds to a rectangular, phase-locked frequency comb with $N$ lines with a spacing of $\Delta f$.

A general schematic of OTDM data transmission with NSS is depicted in Fig. 1. The sinc sequences generated are split into $N$ different branches, where $N$ is again the number of lines in the frequency comb. The next pulse sequence must fall into the zero crossings of the previous to satisfy the orthogonality. Therefore, in each branch the sequence is delayed by a time shift equal to $k/N\Delta f$ with $k = 0, 1, 2, \ldots (N-1)$ as an integer. The pulses are then modulated with the data in each branch and afterwards are multiplexed together in order to have a data rate of $N$ times the data rate of the single TDM channel.

Within the receiver the signal is de-multiplexed and each channel is detected with a coherent receiver, where it is multiplied with an appropriate reference NSS signal. Due to the use of orthogonality the original data can be extracted.

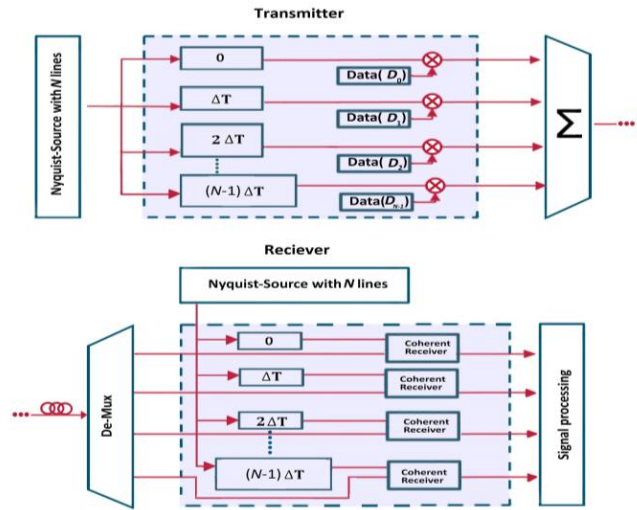

Fig.1 Basic principle of the transmitter and receiver for a data transmission with Nyquist sinc sequences.

The basic advantage of using NSS is that these multiple TDM channels all together have a rectangular bandwidth. These channels can then be multiplexed in the wavelength or frequency domain to form WDM channels without a guardband, that this can be used for enabling high bitrate communications [11].

In order to higher transmission capacity, increasing the bandwidth and/or signal to noise ratio are two main factors [12]. However for limited available bandwidth, higher signal to noise ratio takes more cost and energy. In this respect, complementary Nyquist sinc sequence (CNSS) shows that with two sampled values with less amplitude in comparison with NSS with one sampled value for the same bandwidth and signal to noise ratio has a better performance for high speed data transmission.

In addition to important role of NSS for data transmission, optical delay lines play a key role in optical time division multiplexing (OTDM) system [13]. OTDM system is split to branches with a different time delay in each branch. Different optical propagation path of branches is a very efficient way for time delay. The popular methods are optical fibre and silica spiral, however; it needs larger space for the longer time delay [14-15]. In other side, the precision and stability of the delay is limited.

Alongside, introducing the complementary Nyquist sinc-sequence as a new replacement for data transmission, a possibility of optical time delay by electrical phase shifter for an application in OTDM without requiring any optical delay lines is taken into account.

## II. THEORY

NSS and optical time delays are two important parts of OTDM. In section A, CNNS is defined and then a theory is made for time delay by phase shifting for NSS.

### A. *Complementary Nyquist Sinc Sequences*

The Nyquist source is the first step for OTDM. Since the NSS is an even periodic function, it can be written as a cosine Fourier series for an odd number of *N* as:

$$\frac{\sin(N\pi\Delta ft)}{N\sin(\pi\Delta ft)} = \frac{1}{N} + \frac{2}{N}\sum_{1}^{\frac{N-1}{2}}\cos(2\pi n\Delta ft) \qquad (3)$$

and for an even number of *N* as:

$$\frac{\sin(N\pi\Delta ft)}{N\sin(\pi\Delta ft)} = \frac{2}{N}\sum_{1}^{\frac{N}{2}}\cos(2\pi(n-0.5)\Delta ft). \qquad (4)$$

Thus, NSS is a summation of cosine functions or *N* delta functions equally spaced by $\Delta f$ in the frequency domain, i.e. a rectangular frequency comb.

An arbitrary periodic signal can be represented by a Fourier series:

$$s(t) = a_0 + \sum_{n=1}^{+\infty} a_n \cos(2\pi n\Delta ft) + \sum_{n=1}^{+\infty} b_n \sin(2\pi n\Delta ft) \qquad (5)$$

Accordingly, there is a complementary function of NSS which can represents by sinus Fourier series. In this regard, a second option for data transmission can be presented by complementary signal of NSS. Since the sinus Fourier series is an odd function, CNSS can have with N equal to even numbers. Therefore, leading to:

$$CNSS \stackrel{\Delta}{=} \frac{2}{N}\sum_{1}^{\frac{N}{2}}\sin(2\pi(n-0.5)\Delta ft) \qquad (6)$$

Figure 2 depicts NSS and CNSS for an even number of *N* equal to 4 and 8 with a repetition rate of 10 GHz. CNSS has all properties of NSS, such as orthogonality in one time interval and the same rectangular spectrum. In addition, as can be seen from Fig. 2 that CNSS has two peak points in each time interval with an amplitude less than one. Thus the effect of noise low power as a random phenomenon on the sampled signal statistically is lower in comparison with NSS. Thus, it is a good option for new replacement data transmission for higher signal to noise ratio to achieve a small bit error rate. For instance, for N=4, the effect of noise for two measurements with an amplitude of 0.8 is less than for one measurement with an amplitude of 1 for low power noise.

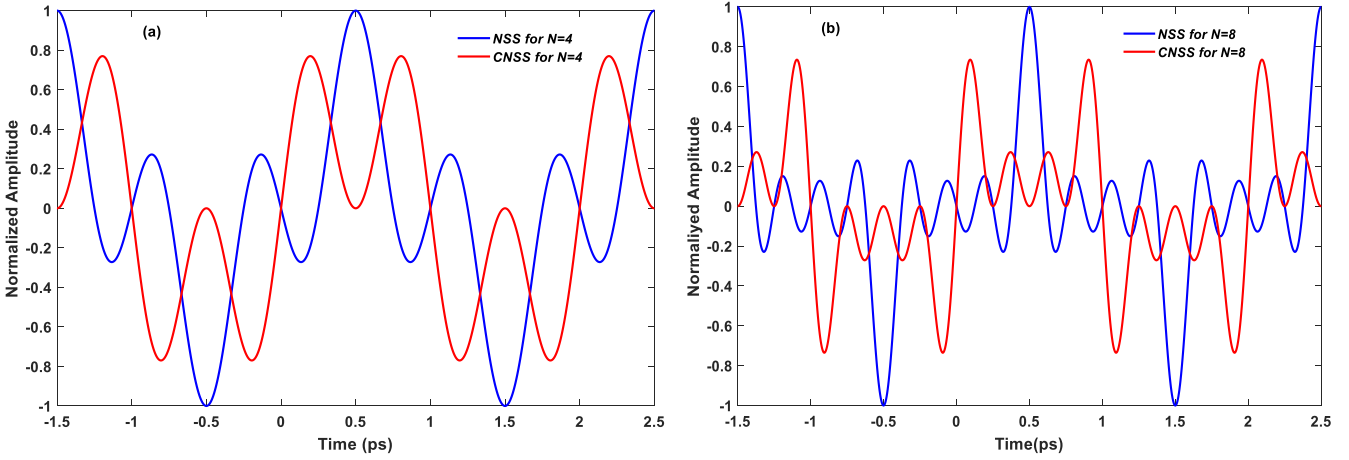

Fig.2 Nyquist sinc sequence (blue) and Complemantary Nyquist sinc sequence (red) for $N = 4$ (a) and $N = 8$ (b) with a repetition rate of 10 GHz in time domain.

### B. *Time Delay by Phase Shifting*

After splitting Nyquist sequence into N branch, signal is delayed correspond to zero crossing time and then data is modulated, as shown in Fig. 1. If time delay applies in Eq. 3, result can be seen in Eq. 7 as a phase shift in cosine function argument. Consequently relation 8 shows the value of shifted phase. Here we continue with N equal to odd, however; method is the same for even number of N.

$$\frac{1}{N} + \frac{2}{N}\sum_{0}^{\frac{N-1}{2}}\cos(2\pi n\Delta f(t-t_0)) = \frac{1}{N} + \frac{2}{N}\sum_{0}^{\frac{N-1}{2}}\cos(2\pi n\Delta ft - 2\pi n\Delta ft_0) \qquad (7)$$

$$\varphi_n = 2\pi n\Delta ft_0 \qquad (8)$$

Where $\varphi_n$ is a phase shift. For Nyquist TDM, $t_0$ must be equal to time value of zero crossing of Nyquist pulses. It is an integer number of $k/N\Delta f$. k corresponds to the number of zero crossing, as well as to the number of branches in OTDM, and continue to $N-1$. By substituting of $k/N\Delta f$ to relation 8, $\varphi_n$ is equal to $\frac{2\pi n}{N}k$.

It is easy to understand the relation between time delay and phase shift in frequency domain.
Relation 9 shows that time delay lead to phase shift in frequency domain.

$$\frac{1}{N} + \frac{2}{N}\sum_{0}^{\frac{N-1}{2}}\cos(2\pi n\Delta ft - 2\pi n\Delta ft_0) \xrightarrow{\mathfrak{I}} \frac{1}{N}\sum_{-\frac{N-1}{2}}^{\frac{N-1}{2}}e^{i2\pi n\Delta ft_0}\delta(f-n\Delta f) \qquad (9)$$

Fig. 3 shows the amplitude and phase frequency of the Nyquist pulses. It can be observed that the spectrum of the NSS consists of N lines.

It is also the same for delayed time signal, however the difference is in the frequency of the phase response.

Time delay lead to change of slope of linear phase. Red and blue lines in Fig. 3(b) illustrate a time delay that corresponds to the first and second zero crossing of the Nyquist pulses, respectively. By changing the slope of the linear phase response, other time delays can be realized easily.

Presented method make a possibility of optical time delay in RF domain by shifting the phase.

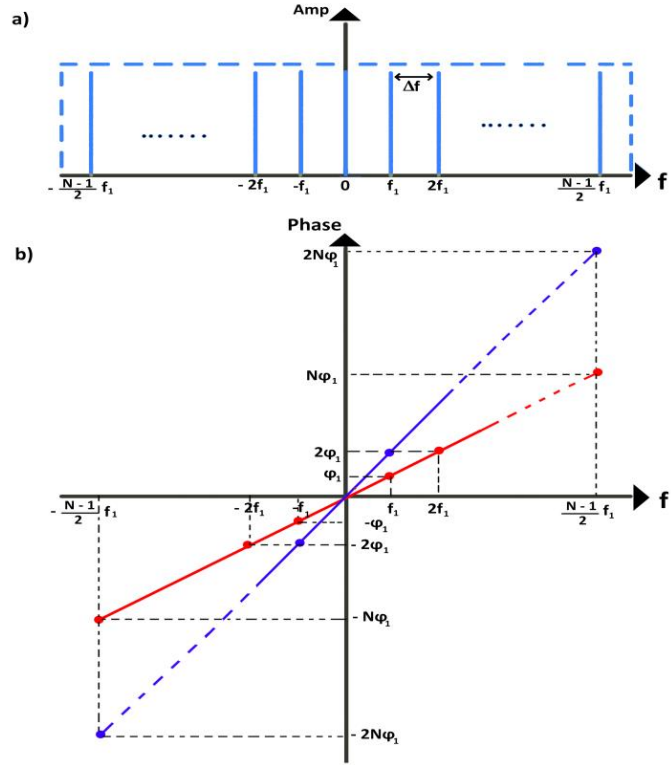

Fig.3. Spectrum of Nyquist sequence: a) Amplitude b) Phase: red and blue lines are corresponding phase for time delay of first and second zero crossing.

### III. RESULTS

Simulation results of BER performance is presented in the first section and then in the next section experimental setup and simulation results for optical time delay by electrical phase shifter for NSS is demonstrated.

#### A. *BER Analysis*

To verify the feasibility of CNSS for OTDM transmission, the BER analysis is carried out by MATLAB Simulink with the schematic setup shown in Fig. 1.

The calculated BER versus signal-to-noise ratio (Eb/No) of NSS and CNSS with $N = 4$ for on/off keying modulation with $2^{13}$ PRBS is depicted in Fig 4. It can be seen that the performance of CNSS is better in comparison with NSS for higher signal to noise ratio. It can be observed that for low Eb/No the difference in not much and is very small, however; after a certain value of Eb/No, here 6.5 dB, the BER plot of CNSS fall down of NSS. The final simulation and experimental results will be added in the future.

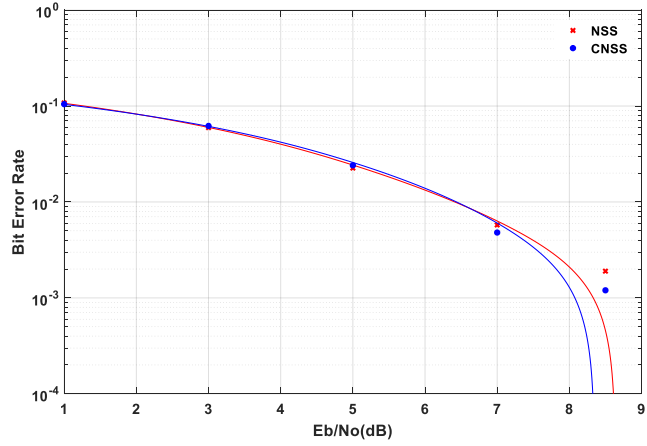

Fig. 4. BER versus Eb/No of a NSS (red) and NNSS (blue) for $N = 4$.

## B. *Optical Time Delay with Electrical Phase Shifter*

The basic setup for the Nyquist pulses with a time delay is shown in Fig. 5. In this experiment, Nyquist source is generated by summation of 3 cosine function with harmonics of $f_1$, $2f_1$ and $3f_1$ plus a DC of $\frac{1}{N}$ volt shown in Eq. 3. where N is equal to 7 and repetition time of $1/f_1 = 1/\Delta f$ by arbitrary wave generator (AWG).

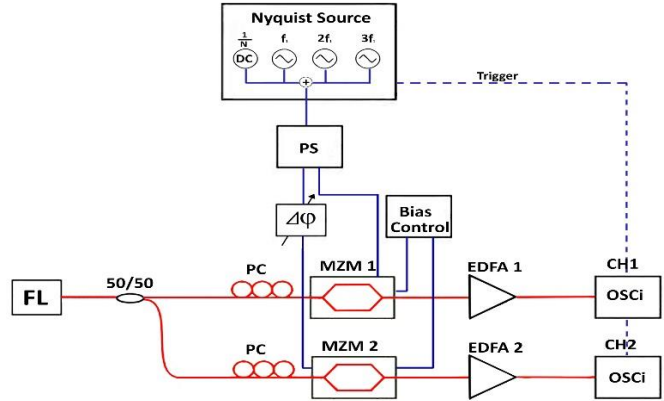

Fig. 5 Experimental setup. FL: fiber laser, PC: polarization controller, PS: Power Splitter, MZM: Mach-Zehnder modulator, EDFA: Erbium doped fibre amplifier, OSCi: Oscilloscope. Blue and red line in the setup show electrical and optical domain respectively.

Then, power divider is split Nyquist sequence into two branch. In this experiment two branches is used instead of 7 branches. However, other branches is made by variable phase shifter and separately data is collected from Ch2 of Oscilloscope. Time delay is done according to Eq. 8 by shifting the phase correspond to $\frac{2n\pi}{7}k$. Ideally phase shifter has a linear variation of phase with frequency. It must be mentioned that phase response of phase shifter must be follow the phase response of Nyquist pulses shown in Fig. 3(b).

The optical carrier frequency is generated by a CW fiber laser (FL) at a wavelength of 1550 nm and with an output power around of 20 dBm. It is split into two branches by a 3dB coupler and is then forwarded into two Mach-Zehnder modulators (MZM). Within the RF section, the 3 frequencies are split into two branches, where an electrical phase shifter is applied to one branch. The electrical signal is transferred to the optical domain by the MZM, resulting in a frequency comb consisting of 7 lines. In order to provide enough power for the detector, the signals are amplified by an Erbium-doped fibre amplifier (EDFA). The bias voltage of MZM is used to control stabilization of the Nyquist sequence. The aligning of polarization controller (PC) is done for maximum transmission.

The experimental results will be shown based on the experiment setup in the future.

# Part II. Complementary Fractional Dimensional Order of Nyquist Sinc Sequences

## I. INTRODUCTION

Recently, beside the integer derivative fractional (non-integer) derivatives and integrals play an important role in theory and applications. The idea of the fractional calculus was planted over 300 years ago in the letters between Leibniz and L'Hospital [16]-[18]. In 1823, Abel investigated the generalized tautochrone problem, and he was the pioneer to apply fractional calculus techniques in a physical problem [16]. Later, Liouville has applied fractional calculus to solve problems in potential theory [18]. Since then, the fractional calculus has triggered the attention of many researchers in all areas of sciences such as fluid mechanics, biology, physics and engineering [19]-[22]. Several attempts have been made to improve the fractional calculus in many different forms of fractional operators [23]-[28] and the solutions of fractional diffrentioal and integral equations such as homotopy perturbation method [29]. However, to the best of my knowledge, beside to need a more deeply research investigation of fractional dimension, a method by the use of fractional dimensions can be required for signal processing and communication system. Here is tried to

## II. REPRESENTATION OF ARBITRARY FUNCTION BY TANGENT LINES IN FRACTIONAL DIMENSIONS

It has shown in [16] that two arbitrary points of function, here $(t_0, f(t_0)) \& (t_n, f(t_n))$, can be connected by a tangent line of function in fractional dimension as shown in Eq. 10.

$$f(t_n) = \left(D^{\alpha_n} f(t)|_{t=t_0}\right)(t_n - t_0) + f(t_0) \tag{10}$$

Where $D^{\alpha_n} f(t)$ is fractional order derivative of the function presented on the following:

$$D_a^{\alpha} f(t) = \frac{d}{dt}\int_a^t \frac{(t-\tau)^{-\alpha}}{\Gamma(1-\alpha)} f(\tau)d\tau = \frac{d}{dt} I_a^{1-\alpha} f(t) \tag{11}$$

Thus, Eq. 10 can be generalized for any arbitrary functions as follows:

$$f(t) = \left(D^{\alpha(t)} f(t)|_{t=t_0}\right)(t - t_0) + f(t_0) \tag{12}$$

It shows that with an appropriate trajectory in fractional dimension can be realized any arbitrary function. This results helps us to make a fractional dimensional signal. For instance, sinus function at $t_0 = 2\pi$ follows as:

$$\sin(t - 2\pi) = \sin(t_0 - 2\pi + \frac{\pi\alpha(t)}{2})(t - 2\pi) = \sin(\frac{\pi\alpha(t)}{2})(t - 2\pi) \tag{13}$$

Thus, sinc function can be produced by sinus function with a trajectory in fractional dimensions.

$$\sin(t - 2\pi) = \sin(\frac{\pi\alpha(t)}{2})(t - 2\pi) \Rightarrow \text{sinc}(t) = \sin(\frac{\pi\alpha(t+2\pi)}{2}) \tag{14}$$

Fractional dimensional trajectory must satisfy the orthogonality condition of sinc function as follows:

$$\int_{-\infty}^{+\infty} (\text{sinc}(t-i)) \times \left(\sin(\frac{\pi\alpha(t)}{2})\right) dt = \begin{cases} 1 & i = 0 \\ 0 & i \neq 0 \end{cases} \tag{15}$$

Where i is an integer number ($i \in \mathbb{N}$).

## III. DIMENSIONAL TRANSFORMATION: DIMENSION ITSELF IS A DIMENSION

Transformation of function from one space to another space is a promising method to reach an easier analysis and more information in different view. In principle, it is a mapping from real world to other virtual world. For instance, Fourier transformation is a mapping from this real world to inverse world, time to frequency or position to momentum. However, if considered Fourier transformation brings a function from time domain to frequency domain, time and frequency as a dimension, however; intrinsically dimension is inside time or frequency. In other words, when we talk about time, automatically it is considered as a dimension. In this point of view, brings out this concept that time is one domain and dimension also another domain which has been coupled together. Thus, it is mentioned that dimension itself is a dimension. In other words, when analyzed the function of f(t) means $f(t, \alpha = 0)$ in an original dimension of function.

In this regard, dimensional transformation is defined as shown in Eq. 16 the variation of function in fractional dimension of $\alpha$.

$$F(\alpha) = \int_{-\infty}^{+\infty} D^\alpha f(t) dt \tag{16}$$

As an example, the dimensional transformation of cosine function is expressed as follows:

$$F(\alpha) = \int_{-\infty}^{+\infty} D^\alpha \cos(\omega t) \, dt = \int_{-T/2}^{+T/2} \omega^\alpha \cos(\omega t + \frac{\pi \alpha}{2}) \, dt = \omega^\alpha \sin(\frac{\pi \alpha}{2}) \tag{17}$$

It can be seen that dimensional transformation can be connected to Mellin transformation. In this regard, Mellin transformation of cosine function is shown in Eq. 18. However, it needs further investigations [17].

$$M\{\cos(\omega t)\}(\alpha) = \Gamma(\alpha) \sin(\frac{\pi \alpha}{2}) \tag{18}$$

Dimensional and inverse dimensional transformations are under further investigations. In the future, we will add the complete result.

## IV. FRACTIONAL DIMENSIONAL ORDER OF NYQUIST SINC SEQUENCES

Fractional dimension order of Nyquist sinc sequence (FDONSS) is defined a variation of function in fractional dimensions with trajectory of $\alpha(t)$ which must satisfy the orthogonality condition. Eq.19 shows the FDONSS for even number of NSS.

$$FDONSS \stackrel{\Delta}{=} D^{\alpha(t)}(NSS) = D^{\alpha(t)}\left(\frac{2}{N}\sum_{n=1}^{N/2} \cos((n-0.5)\omega_0 t)\right)_{t=t_0} = \left(\frac{2}{N}\sum_{n=1}^{N/2} ((n-0.5)\omega_0)^{\alpha(t)} \cos((n-0.5)\omega_0 t_0 + \frac{\pi \alpha(t)}{2})\right) \tag{19}$$

The orthogonality condition is presented in Eq. 20 where $\delta_{ij}$ is Kronecker delta function.

$$\int_{-\infty}^{+\infty} \left(\frac{2}{N}\sum_{n=1}^{N/2} \cos((n-0.5)\omega_0 t - \frac{i\pi}{N})\right) \times \left(\frac{2}{N}\sum_{n=1}^{N/2} ((n-0.5)\omega_0)^{\alpha(t)} \cos((n-0.5)\omega_0 t - \frac{j\pi}{N} + \frac{\pi \alpha(t)}{2})\right) dt = \delta_{ij} = \begin{cases} 1 & i = j \\ 0 & i \neq j \end{cases} \tag{20}$$

In communication system, a method to reach higher data rate is promising. Thus, the complementary-FDONSS (CFDONSS) is defined to reach higher spectral efficiency. The differences of CFDONSS in comparison with FDONSS are in trajectory of $\alpha(t)$ and orthogonality condition.

$$\int_{-\infty}^{+\infty}\left(\frac{2}{N}\sum_{n=1}^{N/2}\cos((n-0.5)\omega_0 t - \frac{i\pi}{N})\right) \times \left(\frac{2}{N}\sum_{n=1}^{N/2}((n-0.5)\omega_0)^{\alpha(t)}\cos((n-0.5)\omega_0 t - \frac{j\pi}{N} + \frac{\pi\alpha(t)}{2})\right)dt = 0 \tag{21}$$

$$\int_{-\infty}^{+\infty}\left(\frac{2}{N}\sum_{n=1}^{N/2}((n-0.5)\omega_0)^{\alpha(t)}\cos((n-0.5)\omega_0 t - \frac{i\pi}{N} + \frac{\pi\alpha(t)}{2})\right) \times \left(\frac{2}{N}\sum_{n=1}^{N/2}((n-0.5)\omega_0)^{\alpha(t)}\cos((n-0.5)\omega_0 t - \frac{j\pi}{N} + \frac{\pi\alpha(t)}{2})\right)dt = \delta_{ij} \tag{22}$$

It is observed that there is no a unique solution for $\alpha(t)$, thus it is tried to find the best solutions of the fractional dimensions trajectory of CFDONSS to be realized better performance for communication system.

Fractional Dimensional trajectory of CFDONSS is under further investigations. In the future, we will add the complete result. To the end, a mixing of conventional OTDM Nyquist block with CFDONSS will be presented for data transmission. The simulation and experimental results will be added in the future.

## IV. CONCLUSION

In Summary, the first, a method for OTDM transmission by CNSS is presented. CNSS, presented by a sinus Fourier series, with taking two sampled values with less amplitude statistically has a better performance of BER for low noise power. Thus for high speed data transmission can be a good replacement compare to NSS. On the other side, a method for tunable Nyquist pulses by a phase change is investigated. Each Nyquist pulses can be represented as a cosine Fourier series. Then, time delay can be realized as a phase shift in argument of cosine functions. Time delay lead to linear phase change in frequency domain. Thus, phase shifter with a linear phase response make a possibility of time delay for Nyquist sequences. The method, theoretically is proved and then experimental setup is demonstrated.

The latter, the basic theoretical framework required to generate a fractional dimension order signal for communication system is introduced. When time is created, automatically dimension is also created as a coupled term. Thus, dimension itself is a dimension. In this regard, dimensional transformation by mapping of function variations in fractional dimension gives an opportunity to extract more information of signal. Furthermore, CFDONSS is defined to reach higher data rate with a combination to conventional OTDM system for data transmission.